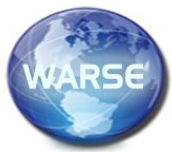

# Analyzing Cyber-Attack Intention for Digital Forensics Using Case-Based Reasoning

**Mohammad Rasmi Al-Mousa**
Department of Software Engineering, Zarqa University, Jordan, mmousa@zu.edu.jo

### ABSTRACT

Cyber-attacks are increasing and varying dramatically day by day. It has become challenging to control cyber-attacks and to identify the perpetrators and their intentions. In general, the analysis of the intentions of cyber-attacks is one of the main challenges in digital forensics. In many cases of cyber-attacks, the analysis of the intent of the attacks determines the strategy and tools used in the attack, thus facilitating the process of identifying the perpetrator of the attack with greater accuracy. In this paper a model will be proposed to analyze the intentions of cyber-attacks. In this proposal, a set of steps will be conceived by linking them with a case-based reasoning methodology. This model will be examined by analyzing the intent of attacks for some cases and comparing the results with other methods of analyzing the intent of attacks. Hopefully the results will determine the intent of cyber-attacks more accurately.

## 1. INTRODUCTION

The Security Report 2019 [1] Study the latest cyber environment risks faced by organizations in the fifth generation. In general, the report concludes that the 2018 attacks had been more prevalent than ever, and are considered to be more focused and stealthy. Such attacks, whether carried out by cybercriminals or nation-states, expose interesting new patterns and motives. Including crypto mining to malware, the vulnerability of mobile devices to threats in the interests of national interests, all have had a significant impact on the risk environment of today. Figure 1 shows the main categories of cyber-attack in the world. Moreover, Table 1 presents the most treatment in the world (2018-2019).

Increased threats when surfing the internet have motivated many researchers and those who are specialized in network security to work together to develop adequate solutions with a view to minimizing network security risks and improving the effectiveness of the network forensics process.

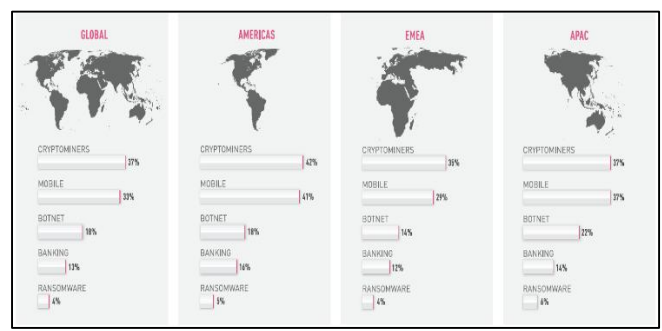

**Figure 1:** Cyber Attack Categories by Region in 2018 [1]

Network forensics is defined, according to Palmer [2] as: "The use of scientifically proven techniques to collect, fuse, identify, examine, correlate, analyze, and document digital evidence from multiple, actively processing and transmitting digital sources for the purpose of uncovering facts related to the planned intent, or measured success of unauthorized activities meant to disrupt, corrupt, and or compromise system components as well as providing information to assist in response to or recovery from these activities".

Modern network forensics models, as mentioned in [3], have several processes embedded in nine phases; planning (preparation), identification (detection), response to accidents, collection, preservation, evaluation (examination), analysis, investigation and presentation.

Network forensics has many research gaps and challenges, such as the variety of data sources, data granularity, data integrity, data as legal evidence, and privacy issues [2-6]. Analysis of Cyber-attack is a major challenge for many people working in network forensics [4].

A useful and accurate analysis of cybercrime leads to the result of the investigation more accurate in addition to saving time and effort at this stage. The process of analyzing cybercrime depends mainly on the accuracy of identifying and collecting evidence associated with it. Where the evidence plays in identifying the perpetrator and the strategy carried out in the attack.

**Table 1:** The most treat in the world (2018-2019) [1]

| Threat Quotation | Source |
|---|---|
| - $2.7 million spent by the City of Atlanta to repair damage from ransomware attack. | Atlanta Journal-Constitution newspaper – www.ajc.com |
| - 76% of organizations experienced a phishing attack in the past year. | 2018 IT Professionals Security Report Survey |
| - The 'AdultSwine' malware was installed up to 7 million times across 60 Children's Games Apps. | Check Point Research Blog |
| - Over 20% of organizations are impacted by Cryptojacking Malware every week. | Check Point ThreatCloud |
| - 40% of organizations were impacted by Cryptominers last year. | Check Point ThreatCloud |
| - The Ramnit Botnet infected 100,000 in just two months. | Check Point Research, Ramnit's Network of Proxy Servers |
| - 49% of organizations experienced a DDoS attack in the past year. | 2018 IT Professionals Security Report Survey |
| - The US and UK formally blamed Russia for the 2017 NotPetya ransomware attack that caused billions of dollars in damages worldwide. | Check Point ThreatCloud |
| - 614 GB of data related to weapons, sensor and communication systems stolen from US Navy contractor, allegedly by Chinese government hackers. | Check Point ThreatCloud |

The evidence plays an important role in analyzing cyber-attack intentions and thus helps investigators improve decision-making and build a coherent case, and therefore for apprehending the perpetrator. Hence, cyber-attack intention analyses support investigators in bringing more successful and accurate criminal cases to a close as mentioned in [7, 8]. Furthermore, it is crucial to accelerate the decision–making processes required for apprehending the perpetrator.

This paper proposes a model to analyze the intentions of cyber-attacks using a case-based reasoning methodology. It will be organized as follows: Section II will present a literature review of related work for cyber-attack intention, and using a case-based reasoning methodology in digital forensics. Section III will propose and describe a proposed model. Section IV describes the experiments. Section V gives analysis and discussion, and Section VI contains the conclusion and further work required.

## 2. BACKGROUND

The United States Department of Justice (DOJ) has defined computer crime as "any violation of criminal law that involves knowledge of computer technology for their perpetration, investigation, or prosecution" (Parker et al., 1989)[9].

Panda Security [10] defined a cybercrime as "a crime where a computer is the object of the crime or is used as a tool to commit an offense. A cybercriminal may use a device to access a user's personal information, confidential business information, government information, or disable a device. It is also a cybercrime to sell or elicit the above information online". However, it classified into two categories; the first one is Crimes that target networks or devices such as malware, viruses, and DoS attacks. Another category is Crimes using devices to participate in criminal activities such as Cyberstalking, identity theft, and phishing emails

In general, the cybercrime could fail into one of the three groups [10, 11]; the first one is against the individuals like harassment via electronic mails, dissemination of obscene material, cyber-stalking, defamation, indecent exposure, cheating, unauthorized control, email spoofing, and fraud. The second group against organization or governments like unauthorized access, cyber terrorism, possession of unauthorized information, distribution of pirated software. The third one is against property like computer vandalism, transmitting virus, netrepass, unauthorized access, intellectual property crimes, and internet thefts. The most dangerous of the cybercrimes which those against society like child pornography, indecent exposure of polluting the youth financial crimes, sale of illegal articles, trafficking, forgery, online gambling.

Digital forensics is concerned with uncovering the facts about cybercrime and solving it. Therefore, it identifies, collects and analyzes the evidence for use in the investigation phase. Therefore, the investigation phase is considered to be costly in terms of time and effort. Thus, the analysis of the evidence of cybercrime with high professionalism and accuracy can reduce the cost of the investigation phase [16-18].

Attack intention, as described in [8], provides investigators with consistent analytical data to generate accurate decision-making. Attack analysis has emerged as a detailed intention study, as discussed in [17], where attacker states and device states are combined to create a list of intentions.

The AIA (Attack Intention Analysis) algorithm, as proposed by [8, 12] aims to predict the attack intention using a mathematical theory called Dumpster–Shafer (D-S), with a probabilistic technique through a causal network as shown in Figure 2.

Case-based reasoning is a computational model consisting of a series of processes. It originates from artificial intelligence and cognitive science that is interested in solving a human problem [13]. It solves and understands new cases through set of processes in a cycle, as shown in Figure 3 by creating a reasoning pattern based on the solutions given in previous

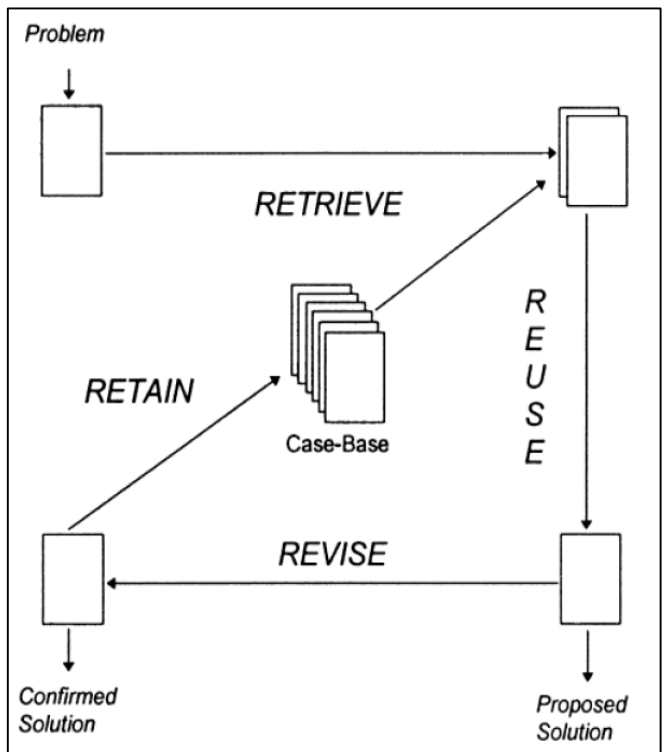

similar cases and reusing information from those cases [14]

**Figure 2:** The AIA (Attack Intention Analysis) algorithm [12]

**Figure 3:** The life cycle of case-based reasoning [14]

## 3. PROPOSED MODEL

In this section, the cyber-attack intention proposed based on the case-based reasoning technique. The proposed model depends on the evidence of the cyber-attack such as exploit ports, implementation of specific functions, using tools and commands, destination and source IP address, type of vulnerability and used protocols. The proposed model as illustrated in Figure 4 includes five processes.

The first process receives all related evidences of the cyber-attack which transferred to the cyber-attack intention (CAI) precedence repository. This repository stores all the cyber-attack intentions associated with it evidences form the previous cyber-attacks. The CAI in this repository analyzed and collected from predefined methods and techniques such as attack intention algorithm (AIA) as presented in [12].

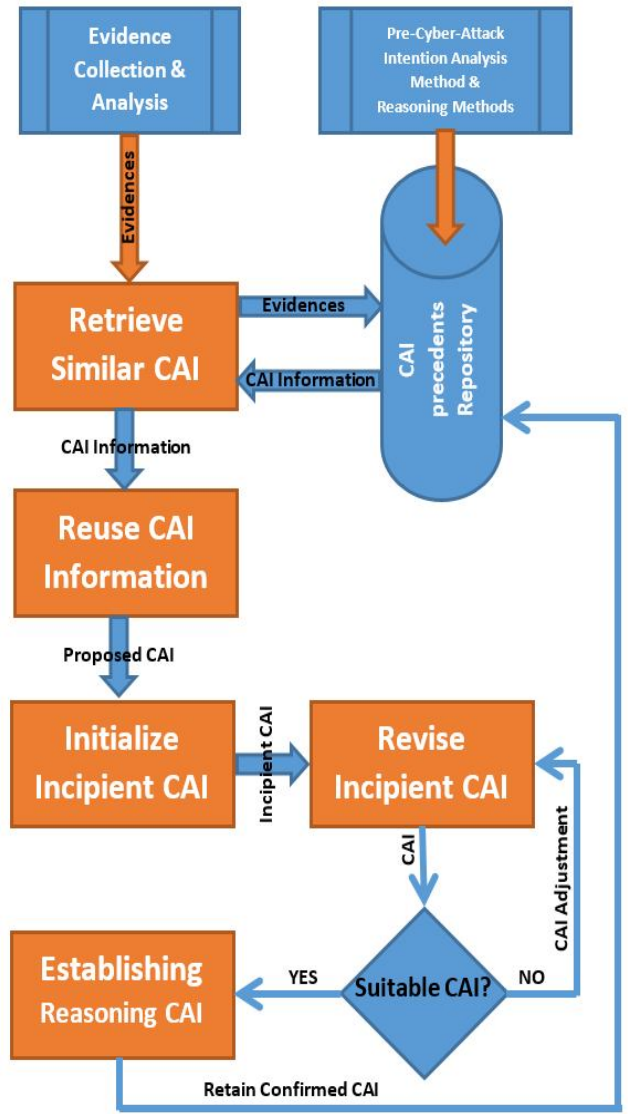

**Figure 4:** The CBR Based Model for Cyber-Attack Intention

The repository correlates the evidences from this process and send the corresponding intention information about the new cyber-attack such as the evidences and weight for each evidence. The process retrieves this information in order to calculate the similarity of the new CAI to a CAI preceded based on the evidences of each one. This process based on the nearest neighbor as presented in [14] to retrieve the similar

CAI. The process weighted of the similarity between the corresponding evidence of the CAI as the following equation:

$$Sim(nCAI, pCAI) = \sum_{ev=1}^{x}(Sim_{ev}(n_{ev}, p_{ev}) * W_{ev}) \quad (1)$$

where *nCAI* is the new cyber-attack intention, *pCAI* is the precedence of the cyber-attack intention, *x* is the number of evidence in each cyber-attack, *i* is an individual evidence from *1* to *x*, $Sim_{ev}$ is a similarity function for evidence *i* in *nCAI* and *pCAI*, and *W* is the weight of evidence which conducted from the retrieved information from the repository.

The similarity of the cyber-attack intention with related information proceeded to the next process which reuses this information in order to produce the proposed intention for the new cyber-attack. The proposed cyber-attack intention entered to the next process and produced the incipient intention for the new cyber-attack. That shall be formally prepared and produced inappropriate coordination and form so that it is understandable and coherent to present it clearly to the investigative phase.

The incipient cyber-attack intention is assessed by linking the intent to the evidence and examining it in accordance with the type of crime and the damage resulting therefrom and then it can be judged whether this intention is appropriate or not. Where it is not appropriate, it is a re-evaluation of the work and the appropriate adjustment or not to adopt this intention. If appropriate, it moves to the next stage in which it takes place to establish the reasoning cyber-attack intention.

Finally, the confirmed reasoning cyber-attack intention will be returned to the repository to be cyber-attack intention precedent in advance to use them to find intentions for new cyber-attack.

## 4. EXPERIMENTAL RESULTS AND ANALYSIS

The experiment in this paper based on the Keylogging cyber-attack, which detected from Avast and Kaspersky antiviruses. Keylogging defined as "a technique that hackers use to copy (or log) the keystrokes of the user and thus, extracting bank account details and other sensitive data." [19].

Keylogging is type of the Botent attack as mentioned in [20], which could be a Peer-to-peer model or Client-server model. Based on [12], there is a lot of Botnet intentions such as stealing secret data and exposing the victim's compassionate information, gathering all kinds of information for his nefarious purposes; using it to spy on users of compromised computers; watching everything the victim does. Moreover, each of these intentions is linked to a set of evidence.

Table 2, presents some evidence related to the Keylogging cyber-attack. The weight for each evidence will be retrieved from the cyber-attack intention (CAI) precedence repository as one part of CAI information.

**Table 2:** Some evidences for Keylogging cyber-attack

| Evidence Number | Evidence |
|---|---|
| ev01 | Exploit the windows registry and read from the victim's device invalid registry entries. |
| ev02 | Implements a weedfind feature that can be used for information retrieval. Or use the bot of reverb. |
| ev03 | Implement several functions to recover the compromised machine's registered owner and company. |
| ev04 | Using W32/Agobot |
| ev05 | Using commands such as sysinfo and netinfo. |

Based on the some but not all Botnets attacks intention as detected in [12], the similarity of the new cyber –attack (nCAI) (Keylogging) to cyber-attack preceded (pCAI) are presented in Table 3. This weighted of the similarity between the corresponding evidence of the new cyber –attack (Keylogging).

**Table 3:** Similarity of (nCAI) (Keylogging) to (pCAI)

| Intention Number | pCAI [12] | Sim(nCAI,pCAI) |
|---|---|---|
| 1. | To collect all kinds of information for his nefarious purposes. | 0.85 |
| 2. | To be employed to spy on the users of the compromised machines. (spying) | 0.79 |
| 3. | To observe everything the victim is doing, Key logging, stealing secret data or to reveal very sensitive information on the victim. | 0.91 |
| 4. | To gain financial advantages. (Installing advertisement add-ons and Browser Helper Objects (BHOs)) | 0.68 |
| 5. | To grab e-mail addresses or other contact information from the compromised machine. | 0.64 |
| 6. | To steal CD-keys from the victim's hard disk or any software has been legally purchased. | 0.62 |
| 7. | To gets an overview of the hardware configuration or retrieving information about the victim's host.( the speed of the CPU, the uptime, and IP address) | 0.67 |
| 8. | Attacker plans to sell or rent his bots to others. | 0.53 |
| 9. | To spread further, usually by automatically scanning whole network ranges and propagating themselves via vulnerabilities. | 0.43 |
| 10. | To search the hard drive of all victims for sensitive files based on a regular expression. | 0.71 |
| 11. | To spreading new malware. | 0.38 |

Figure 5 shows and builds a preliminary concept in analyzing and determining the intent of a cyber-attack so as to minimize the probability of investigators as well as make them concentrate in intentions with high proportions in similarity with the previous one.

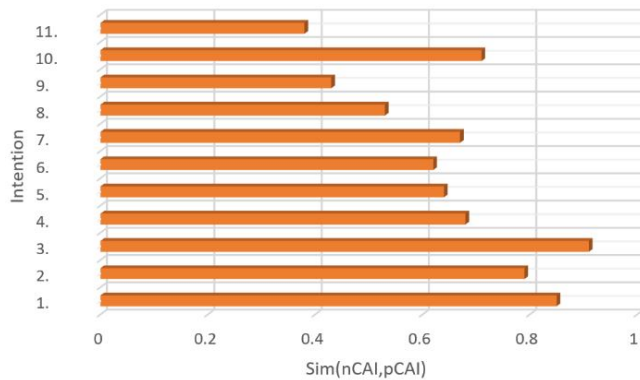

**Figure 5:** Similarity Overview of (nCAI) (Keylogging) to (pCAI)

## 5. CONCLUSION AND FUTURE WORK

In this paper a model to analyze the intentions of cyber-attacks proposed using a case-based reasoning computation technique. The proposed model depends on the evidence of the cyber-attack and has five processes includes retrieving information from the cyber-attack intention (CAI) precedence repository to obtain a similar cyber-attack intention. Then reusing this information to propose a CAI then initializing the incipient CAI. After that revising the incipient CAI to check if it suitable to be an intention for the new cybercrime. If it is then the reasoning CAI established and retained to the cyber-attack intention (CAI) precedence repository to used again to analyzing a new cyber-attack intention. The results construct a preliminary view to evaluate and assess the intent of a cyber-attack in order to reduce the risk of investigators and to focus them on actions of large proportions similar to the previous one.

This research motivates a network forensics researcher to propose new methods and techniques to improve the phases of network forensics, especial in analysis and investigation phases.

## ACKNOWLEDGEMENT

This research is funded by the Deanship of Research and Graduate Studies at Zarqa University /Jordan.